\documentclass[aps,pre,amsmath,amssymb,lengthcheck,showpacs,superscriptaddress]{revtex4-1}
\usepackage{graphicx}
\usepackage{color}
\usepackage{amsfonts}
\usepackage{physics}
\usepackage{braket}
\usepackage{multibib}
\usepackage[english]{babel}

\begin{document}


\title{Low-frequency vibrations of jammed packings in large spatial dimensions}

\author{Masanari Shimada}

\email{masanari-shimada444@g.ecc.u-tokyo.ac.jp}

\affiliation{Graduate School of Arts and Sciences, The University of Tokyo, Tokyo 153-8902, Japan}

\author{Hideyuki Mizuno}

\affiliation{Graduate School of Arts and Sciences, The University of Tokyo, Tokyo 153-8902, Japan}

\author{Ludovic Berthier}

\affiliation{Laboratoire Charles Coulomb (L2C), Universit\'e de Montpellier, CNRS, 34095 Montpellier, France}

\author{Atsushi Ikeda}

\affiliation{Graduate School of Arts and Sciences, The University of Tokyo, Tokyo 153-8902, Japan}
\affiliation{Research Center for Complex Systems Biology, Universal Biology Institute, University of Tokyo, Komaba, Tokyo 153-8902, Japan}

\date{\today}


\begin{abstract}
Amorphous packings prepared in the vicinity of the jamming transition play a central role in theoretical studies of the vibrational spectrum of glasses.
Two mean-field theories predict that the vibrational density of states $g(\omega)$ obeys a characteristic power law, $g(\omega)\sim\omega^2$,
called the non-Debye scaling in the low-frequency region.
Numerical studies have however reported that this scaling breaks down at low frequencies, due to finite dimensional effects. In this study, we prepare amorphous packings of up to $128000$ particles in spatial dimensions from $d=3$ to $d=9$ to characterise the range of validity of the non-Debye scaling. Our numerical results suggest that the non-Debye scaling is obeyed down to a frequency that gradually decreases as $d$ increases, and possibly vanishes for large $d$, in agreement with mean-field predictions. We also show that the prestress is an efficient control parameter to quantitatively compare packings across different spatial dimensions. 
\end{abstract}


\maketitle

\section{Introduction}
Amorphous solids represent a ubiquitous state of matter.
Despite their importance, understanding their properties has been a challenge in condensed matter physics for a long time.
However, recent studies on the jamming transition have opened the door to fundamental progress to understanding the physics of glasses~\cite{O'Hern2002Random,O'Hern2003Jamming,Silbert2005Vibrations,Wyart2005Geometric,Wyart2005Effects,Silbert2009Normal,Hecke2010Jamming,Wyart2010Scaling, Degiuli2014Effects, Franz2015Universal}.
When a packing of athermal particles interacting through a repulsive, finite-range potential is compressed, particles start to overlap with each other at a certain density, where the packing acquires a finite pressure $p$ and shear modulus $G$, i.e., it becomes a solid.
This is the jamming transition.
Jammed systems can be considered as a simple model system for glasses, and such models have enabled the construction of sophisticated theories~\cite{Wyart2005Geometric,Wyart2005Effects,Wyart2010Scaling,Degiuli2014Effects,Franz2015Universal} that rely on the specific critical properties of the jamming transition. 

Close to the transition, mechanical and geometrical observables display  power-law dependences on the distance to the jamming transition point~\cite{O'Hern2002Random,O'Hern2003Jamming}. Usually, the pressure $p$ or the excess packing fraction $\Delta\phi = \phi-\phi_J$ are used to measure the distance from the jamming transition, where $\phi_J$ is the packing fraction at the transition point.
Both quantities are easy to control in simulations and experiments, and they obey the simple relation: $p \sim \Delta\phi$.
The shear modulus $G$ follows instead the scaling law $G \sim p^{1/2}$ close to jamming~\cite{O'Hern2002Random,O'Hern2003Jamming}, and thus, the system gradually acquires rigidity as pressure increases above jamming.
The contact number per particle, $z$, characterizes the geometrical properties of such packings, and it becomes exactly twice the spatial dimensionality $d$ at the transition point, which can be understood from the Maxwell criterion~\cite{Maxwell1864}. Defining the excess contact number as $\delta z = z-2d$, the scaling relation $\delta z\sim p^{1/2}$ then holds~\cite{O'Hern2002Random,O'Hern2003Jamming}.
In addition to the above scaling relations, many other quantities show power-law behaviors, such as the radial distribution function and the force distribution~\cite{O'Hern2002Random,O'Hern2003Jamming,Charbonneau2012,Lerner2013,Charbonneau2015Jamming}, for which accurate theoretical descriptions are now available~\cite{Wyart2012,Charbonneau2014b}.

In addition, the vibrational properties of jammed systems have attracted intense attention. One motivation is that they are expected to shed new light on the low-frequency vibrational properties of structural glasses, which govern their low-temperature thermal properties~\cite{Kittel1996Introduction,Zeller1971Thermal,Anderson1972Anomalous}, the structural relaxation of supercooled liquids~\cite{Oligschleger1999,Widmer-Cooper2009}, and their mechanical failure under load~\cite{Maloney2006,Tanguy2010,Manning2011}.
In particular, the vibrational density of states (vDOS) $g(\omega)$, where $\omega$ is the frequency, is a central quantity for characterizing the vibrational properties of a material.
Near the jamming transition point, a characteristic plateau $g(\omega)\sim\omega^0$ is observed~\cite{O'Hern2002Random,O'Hern2003Jamming, Silbert2005Vibrations, Silbert2009Normal}. The onset frequency of the plateau is denoted by $\omega^\ast$, and this onset frequency goes to zero as the system approaches the jamming transition, with a power-law dependence of $\omega^\ast\sim p^{1/2}$~\cite{Silbert2005Vibrations, Silbert2009Normal}.
Below $\omega^\ast$, the vDOS shows a quadratic frequency dependence $g(\omega)\sim (\omega/\omega^\ast)^2$~\cite{Degiuli2014Effects,Charbonneau2016Universal,Mizuno2017Continuum}.
Since this dependence is independent of the number of spatial dimensions $d$, it is different from the Debye law $g_{\mathrm{Debye}}(\omega)\sim\omega^{d-1}$~\cite{Kittel1996Introduction}, except in the important case where $d=3$. Hence, this is called the non-Debye scaling~\cite{Charbonneau2016Universal}.
The non-Debye scaling does not seem to extend to zero frequency. Instead, below a certain frequency, spatially localized vibrations called quasilocalized vibrations (QLVs) coexist with plane waves, or phonons~\cite{Mizuno2017Continuum,Shimada2018Anomalous,Wang2018}.
These QLVs obey a quartic power law~\cite{Lerner2016Statistics,Mizuno2017Continuum,Shimada2018Anomalous,Lerner2017Effect,Shimada2018Spatial,Wang2018,Kapteijns2018Universal}, namely, $g_{\mathrm{QLV}}(\omega)\sim(\omega/\omega^\ast)^4$~\cite{Mizuno2017Continuum,Shimada2018Spatial}.
As a result, particularly in three-dimensional space~($d=3$), the non-Debye scaling manifests itself as a peak in the reduced vDOS $g(\omega)/\omega^2$, called the boson peak~\cite{buchenau_1984,Yamamuro_1996,Kabeya_2016}.
We remark that since the Debye law is $g_{\mathrm{Debye}}(\omega)\sim\omega^{2}$ for $d=3$, the reduced vDOS is $g(\omega)/\omega^2 \sim g(\omega)/g_{\mathrm{Debye}}(\omega)$.

To explain these observations, two kinds of mean field theories have been developed: the replica theory for a perceptron~\cite{Franz2015Universal,Ikeda2018Note,Ikeda2018Universal} and the effective medium theory (EMT)~\cite{Wyart2010Scaling,Degiuli2014Effects,Degiuli2014Force}.
The former~\cite{Franz2015Universal,Ikeda2018Note,Ikeda2018Universal} addresses a perceptron model, which is considered to belong to the same universality class as jammed materials.
The latter~\cite{Wyart2010Scaling,Degiuli2014Effects,Degiuli2014Force} maps a jammed solid onto a disordered lattice and then considers the resulting equations of motion. Both theories predict that the vDOS should become flat for $\omega > \omega^\ast$ and should exhibit the non-Debye scaling $g(\omega) \sim \omega^2$ for $\omega < \omega^\ast$.
Both these theories obtain the non-Debye scaling as a consequence of the marginal stability of the system~\cite{Degiuli2014Effects,Franz2015Universal,Muller2015Marginal}. Mathematically, marginal stability translates into full replica symmetry breaking in the replica theory and an elastic instability in the effective medium theory. In particular, when the system is on the verge of instability, the non-Debye scaling becomes gapless~\cite{Franz2015Universal,Degiuli2014Effects}, namely the scaling $g(\omega) \sim \omega^2$ extends down to zero frequency and should dominate (and replace) the usual Debye law for solids.
These theories therefore predict that for three dimensions, the boson peak will not be a `peak' in a marginally stable glass but that instead, the reduced vDOS $g(\omega)/\omega^2$ should take at low $\omega$ a constant value which is larger than the Debye prediction.

However, simulations of three-dimensional systems have found that the boson peak is, perhaps unsurprisingly given its name, just a peak. The non-Debye scaling does not extend to zero frequency, and instead, QLVs appear at low frequency and dominate the low frequency behaviour.
This discrepancy between the simulations and theories can be attributed to either of two incompatibilities between them.
(1) The theories are of mean-field nature and are expected to work well only in the infinite-dimensional limit, whereas the simulations are performed in a finite number of dimensions (mostly, three). Therefore, the breakdown of the non-Debye scaling and the appearance of the QLVs may be due to finite-dimensional effects.
(2) Theories do not directly address the packing of particles.
The replica theory considers a perceptron model, whereas effective medium theory focuses on a spring network on a disordered lattice.
The real packings of the particles may not be similar to these models, even in the infinite-dimensional limit.

To understand the discrepancy between the simulations and theories, it is necessary to numerically access the full frequency range of the non-Debye scaling for large-dimensional systems, to see whether and how the vDOS converges to the theoretical prediction in the large $d$ limit.
Previously, Charbonneau {\it et al.}~\cite{Charbonneau2016Universal} studied vibrations in packings of particles in $d=3$--$7$ and provided evidence in favor of the existence of a region of quadratic non-Debye scaling in these dimensions. Kapteijns {\it et al.}~\cite{Kapteijns2018Universal} studied $d=2$--$4$ and established instead the existence of the quartic law due to QLVs in these dimensions, using similar models and parameters Charbonneau {\it et al.}. Therefore, the validity range of the non-Debye scaling was not accessed in these earlier studies, and the important question regarding the discrepancy with the theory has not been addressed.

In this work, we study the vibrational properties of packings of up to $N=128000$ particles in dimensions $d=3$--$9$ and answer the questions raised above. Before studying the dependence of the vDOS on the spatial dimensionality, we first consider the appropriate normalization of our control parameter.
Although the excess packing fraction $\Delta\phi$ and the pressure $p$ are useful quantities in low-dimensional systems for characterizing the distance from the jamming transition, their complicated dependence on $d$ makes it difficult to compare different dimensionalities.
Instead, we use the prestress, $e$~\cite{Wyart2005Effects,Degiuli2014Effects}, which we define shortly. This quantity is more easily normalized and handled in different spatial dimensions.
By analysing the excess contact number and the onset frequency for the characteristic plateau in the vDOS, we show that this choice enables us to best compare packings in different dimensions.
We then extract the onset frequency of the non-Debye scaling and study its dependence on the spatial dimensionality.
We find that the onset frequency decreases with increasing $d$, suggesting that the non-Debye scaling region extends to lower frequency with an increasing number of dimensions, at least up to $d=9$.
Our numerical results suggest that the vDOS converges to the predicted form and that the non-Debye scaling becomes gapless even in real particle systems in the infinite-dimensional limit. 


In Sec.~\ref{methods} we present the model and methods used in the present study. 
In Sec.~\ref{results} we present the numerical results, and we discuss them in Sec.~\ref{discussion}.


\section{Model and Methods}

\label{methods}

We generated monodisperse packings of particles of mass $m$ in a periodic cubic box.
The particles interact via a finite-range harmonic potential:
\begin{equation}
    \phi(r) = \frac{\epsilon}{2}\left(1-\frac{r}{\sigma}\right)^2H(\sigma-r),
\end{equation}
where $r$ is the distance between two particles; $\epsilon$ and $\sigma$ are the characteristic energy and length scales, respectively; and $H(x)$ is the Heaviside step function, i.e., $H(x) = 1$ for $x \ge 0$ and $H(x) = 0$ for $x<0$.
In this paper, we report the length, mass, and time in units of $\sigma$, $m$, and $\sqrt{m\sigma^2/\epsilon}$, respectively.
The considered spatial dimensionalities are $d=3$, $4$, $5$, $6$, $7$, $8$, and $9$.
For each $d$, we prepared packings of $N=8000$, $16000$, $32000$, $64000$ and $128000$ particles.
The mechanically stable configurations (inherent structures) were generated via quenching from random configurations using the FIRE algorithm~\cite{Bitzek2006Structural}.

After obtaining the inherent structures and removing rattlers (with a contact number of less than $d$), we carried out a normal mode analysis.
We calculated the dynamical matrix, which is the second derivative of the system potential $U=\sum_{i<j} \phi(r_{ij})$ with respect to the particle positions $\{\boldsymbol{r}_i\}_{i=1}^N$:
\begin{equation}\label{eq:dynamical matrix}
    \begin{split}
    &\frac{\partial^2 U}{\partial\boldsymbol{r}_i\partial\boldsymbol{r}_j} \\
    &= \left\{\hat{\boldsymbol{r}}_{ij}\hat{\boldsymbol{r}}_{ij}^T\phi''(r_{ij}) - (I_d - \hat{\boldsymbol{r}}_{ij}\hat{\boldsymbol{r}}_{ij}^T)\left[-\frac{\phi'(r_{ij})}{r_{ij}}\right]\right\}\delta_{\left<ij\right>} \\
    &+ \sum_{j'(\neq i)}\left\{\hat{\boldsymbol{r}}_{ij'}\hat{\boldsymbol{r}}_{ij'}^T\phi''(r_{ij'}) - (I_d - \hat{\boldsymbol{r}}_{ij'}\hat{\boldsymbol{r}}_{ij'}^T)\left[-\frac{\phi'(r_{ij'})}{r_{ij'}}\right]\right\}\delta_{ij},
    \end{split}
\end{equation}
where $\hat{\boldsymbol{r}}_{ij} = \boldsymbol{r}_{ij}/r_{ij}= (\boldsymbol{r}_{i}-\boldsymbol{r}_{j})/|\boldsymbol{r}_{i}-\boldsymbol{r}_{j}|$ is the unit vector along the direction of the vector joining particles $i$ and $j$, $I_d$ is the $d\times d$ identity matrix, and $\delta_{\left<ij\right>} = 1$ when $i$ and $j$ are in contact.
We calculated all eigenvalues $\lambda^k$ and the associated eigenvectors $\boldsymbol{e}^k = (\boldsymbol{e}^k_1\cdots\boldsymbol{e}^k_N)$ of the dynamical matrix using LAPACK~\cite{lapack} for the configurations with $N=8000$, where the $k=1,\cdots,dN$ are the labels of the eigenmodes.
For $N\geq16000$, it would be practically impossible to obtain all the eigenmodes; therefore, we instead calculated the smallest eigenvalues and the associated eigenvectors using ARPACK~\cite{arpack}.

Based on the vibrational eigenmodes, we define three quantities.
The first is the participation ratio~\cite{Mazzacurati1996Low,Schober2004Size,Taraskin1999Anharmonicity}
\begin{equation}
    P^k = \frac{1}{N} \left[ \sum_{i=1}^N (\boldsymbol{e}_i^k\cdot\boldsymbol{e}_i^k)^2 \right]^{-1}.
\end{equation}
This is a measure of the degree of localization.
When all particles vibrate equally, $P^k=1$, and when only one particle vibrates, $P^k=1/N$.
The other quantities are the vDOS
\begin{equation}
    g(\omega) = \frac{1}{dN-d}\sum_{k=1}^{dN-d}\delta(\omega-\omega^k),
\end{equation}
where $\omega^k = \sqrt{\lambda^k}$ is the eigenfrequency, and the associated cumulative distribution (CD)
\begin{equation}
    C(\omega) = \int_0^{\omega}d\omega' g(\omega') = \frac{1}{dN-d}\sum_{k=1}^{dN-d}H(\omega-\omega^k).
\end{equation}
Note that in the definitions of the vDOS and the CD, we exclude the $d$ zero modes corresponding to global translations.
Although the vDOS is sufficient for the study of the non-Debye scaling region, the CD is free from binning errors and thus is helpful for studying the lowest-frequency regime.
We calculated the vDOS and the CD for each $N$, and the averaged results are presented.
Since we are mainly interested in the non-Debye scaling $g(\omega)\sim\omega^2$, we define and study  the reduced vDOS $\Tilde{g}(\omega) = g(\omega)/\omega^2$ and the reduced CD $\Tilde{C}(\omega) = C(\omega)/\omega^3$
\footnote{
In studies of the boson peak~\cite{buchenau_1984,Yamamuro_1996,Kabeya_2016}, one usually considers another definition of the reduced vDOS, namely, $g(\omega)/g_{\mathrm{Debye}}(\omega)\propto g(\omega)/\omega^{d-1}$, to focus on the modes in excess over the Debye prediction.
For the three-dimensional case~($d=3$), $g(\omega)/g_{\mathrm{Debye}}(\omega)$ is accidentally proportional to $\tilde{g}(\omega) = g(\omega)/\omega^2$.}
.

\section{Results}

\label{results}

\subsection{Scaling based on the prestress}

\begin{figure*}
    \centering
    \includegraphics{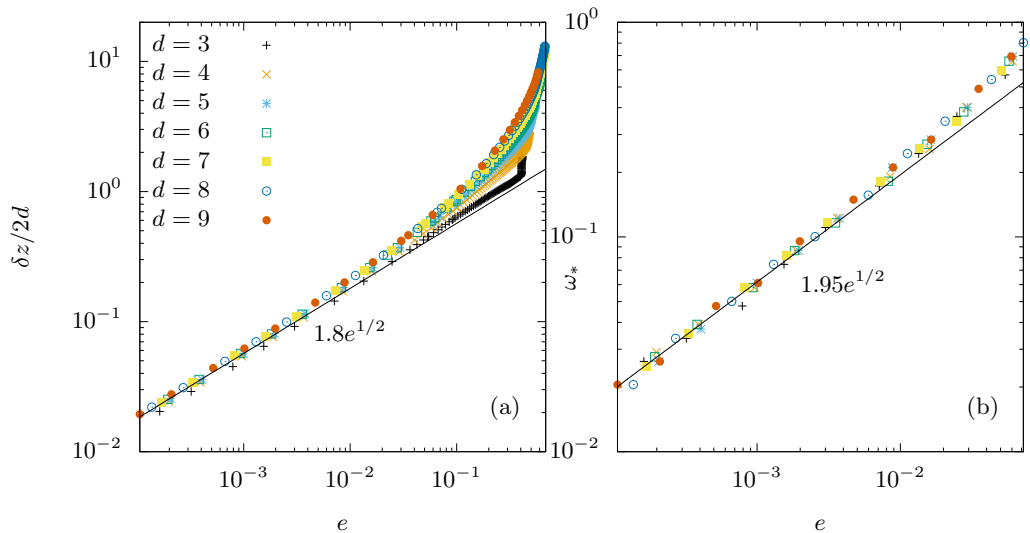}
\caption{
(a) The excess contact number $\delta z$ divided by $2d$ as a function of the prestress $e$.
(b) The onset frequency $\omega^\ast$ as a function of the prestress.
In both plots, the solid lines are proportional to $e^{1/2}$, with the indicated numerical prefactors.
}
    \label{fig:fig1}
\end{figure*}

To discuss the vibrational properties of jammed systems, we first have to specify a control parameter to express the distance from the jamming transition.
Usual choices are the excess packing fraction, $\Delta\phi = \phi - \phi_J$ or $\Delta\phi/\phi_J$, and the pressure, $p$~\cite{O'Hern2002Random,O'Hern2003Jamming,Silbert2005Vibrations,Silbert2009Normal,Hecke2010Jamming,Charbonneau2016Universal}. However, the excess packing fraction is not convenient for our study because $\phi_J$ depends on $d$ and on the preparation protocols in a highly nontrivial manner~\cite{Charbonneau2011,Chauduri2010Jamming}, making it difficult to compare different dimensionalities and different protocols~\footnote{
$\phi_J$ is expected to follow $\phi_J \sim 2^{-d} d$ in large dimensions~\cite{Charbonneau2011}.
Even if this is the case, however, the dimensions in our study ($d \leq 9$) would be outside the scaling region of this relation, and thus, rational control of this quantity would be difficult.
}.
The pressure is also not a very convenient choice because it is equal to the total virial divided by the volume of the system~\cite{Hansen2006Theory}, and the volume of a jammed system also depends on $d$ in a very complicated way.
As a result, the pressure and excess packing fraction cannot be well controlled for different dimensionalities.

A natural choice for the control parameter is found by considering the dynamical matrix given in Eq.~(\ref{eq:dynamical matrix})~\cite{Wyart2005Effects,Degiuli2014Effects,Degiuli2014Force}.
In Eq.~(\ref{eq:dynamical matrix}), there are two terms in curly brackets.
The first one is proportional to the second derivative of the potential $\phi''(r)=1$, and the second one is proportional to the force $-\phi'(r)=1-r$.
The first term is always positive and can be simply interpreted as the stiffness of the spring between $i$ and $j$.
The second term is negative and destabilizes motions along the $d-1$ directions perpendicular to $\hat{\boldsymbol{r}}_{ij}$.
The competition between these two terms is crucial for the stability of the system.
Therefore, we define the ratio of the second term to the first term as $e = (d-1)\left<-\phi'(r_{ij})/r_{ij}\phi''(r_{ij})\right>_{ij} = (d-1)\left<1/r_{ij}-1\right>_{ij}$, where $\left<\bullet\right>_{ij}$ is the average over all contacts.
This ratio is usually called the prestress~\cite{Wyart2005Effects,Degiuli2014Effects,Degiuli2014Force}.
Note that the prestress is proportional to the pressure near the jamming transition with a fixed $d$.
Since EMT predicts that $\omega^\ast\sim\delta z\sim e^{1/2}$~\cite{Wyart2005Effects,Degiuli2014Effects}, the prestress is a more fundamental quantity than the pressure is for discussing the scaling relation.
Furthermore, the excess contact number is of order $d$, and thus, a suitable normalization for it is $\delta z/2d = z/2d-1$~\cite{Wyart2010Scaling,Degiuli2014Effects,Degiuli2014Force}.

Figure~\ref{fig:fig1} shows (a) the excess contact number $\delta z/2d$ and (b) the onset frequency of the plateau in the vDOS $\omega^\ast$ as functions of $e$.
The former is measured for $N=16000$, and the latter is measured for $N=8000$~\footnote{
In practice, $\omega^\ast$ is estimated by comparing the vDOS of the original system with that of the unstressed system.
The unstressed system is the jammed system in which the second, force-proportional term in Eq.~(\ref{eq:dynamical matrix}) is neglected.
We define $\omega^\ast$ as the frequency of the intersection of these two vDOSes.
}.
Since $\omega^\ast$ cannot be defined far from the jamming transition, we do not have data for $e\gtrsim 0.07$ in Fig.~\ref{fig:fig1}(b).
In both Figs.~\ref{fig:fig1}(a) and (b), the data collapse to a single master curve.
For $e \lesssim 0.01$, we obtain $\delta z/2d \approx 1.8 e^{1/2}$ and $\omega^\ast \approx 1.95 e^{1/2}$ (solid lines), which work almost perfectly in all dimensions.
For $e \gtrsim 0.01$,  $\delta z/2d$ and $\omega^\ast$ start to deviate from these power laws.
However, the data still collapse to a single master curve even in this region, especially in large $d$.
This suggests that our normalization is valid even far from the jamming transition point, where these quantities no longer follow a power-law scaling.
At even larger $e$, the excess contact number exhibits a kink: the most visible case is at $e \sim 0.4$ in $d=3$.
This kink corresponds to the crossover to ``deeply jammed'' solids, in which particles interact with their second nearest neighbors~\cite{Zhao2011New,Wang2012Non-monotonic}.

From now on, we use $e$ as the control parameter to discuss the vibrational properties of jammed systems in various spatial dimensions.


\subsection{Vibrational properties}

Having identified the appropriate control parameter $e$, we then generate packings with identical prestress values, $e = 0.25$, $0.2$, $0.15$, $0.1$, $0.05$, and $0.01$ for dimensions $3\leq d\leq 9$ and particle number $8000\leq N\leq 128000$, and analyzed their vibrational properties. 
To generate packings with a given target prestress, we first obtain the mechanically stable configuration at high density by quenching (see Sec.~\ref{methods}), and iteratively compressed or decompressed them until the target prestress is reached~\cite{Goodrich2013}.
Note that for the case $N=128000$, we could prepare only packings of $e\geq0.15$ in $d=6$, $e=0.25$ in $d=7$, $e=0.25$ in $d=8$, and $e=0.2$ in $d=9$.

\begin{figure}[b]
    \centering
    \includegraphics[width=0.45\textwidth]{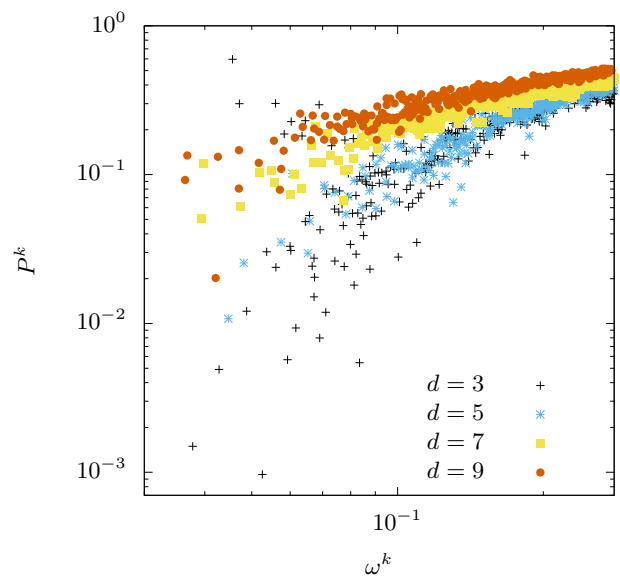}
\caption{
The participation ratio as a function of the frequency for $e=0.1$.
Each point indicates an eigenmode.
We show the data only in odd numbers of dimensions for visualization purposes.
}
    \label{fig:fig2}
\end{figure}

First, we plot the participation ratio $P^k$ in Fig.~\ref{fig:fig2}.
Since the results are qualitatively the same for all $e$, we only show the data for $e=0.1$ in odd numbers of dimensions.
From this figure, we see that the vibrations in the high-frequency regime have large $P^k$ values, i.e., they are extended.
On the other hand, in the low-frequency regime, the participation ratio gradually decreases, signaling the existence of QLVs.
As the dimensionality increases, the vibrations become more extended, and the onset frequency where the vibrations start to be localized decreases, as previously observed in Ref.~\cite{Charbonneau2016Universal}.
This implies that the non-Debye scaling $g(\omega)\sim\omega^2$, which consists of extended vibrations, may be obeyed over a broader range of frequencies towards small frequencies as the number of spatial dimensions increases.

\begin{figure}
    \centering
    \includegraphics[width=0.45\textwidth]{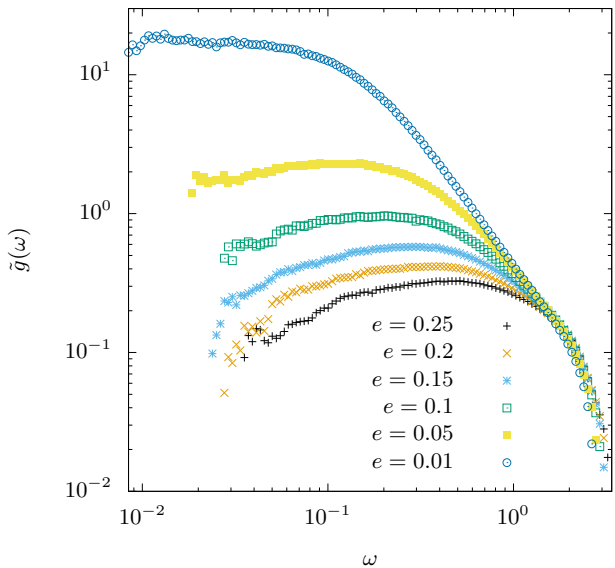}
\caption{
The reduced vDOS in $d=6$ for $e = 0.25$, $0.2$, $0.15$, $0.1$, $0.05$, and $0.01$.
}
    \label{fig:fig3}
\end{figure}

To quantitatively investigate the non-Debye scaling, we calculated the vDOS and the CD.
In Fig.~\ref{fig:fig3}, we plot the reduced vDOS for all $e$ in $d=6$.
In all cases, we observe the plateaus $\Tilde{g}(\omega)\sim\omega^0$, which correspond to the non-Debye scaling $g(\omega) \sim \omega^2$.
Interestingly, although the non-Debye scaling was initially discussed in the context of the jamming transition, it can be observed even for $e \gtrsim 0.01$, where the power-law relation between the excess contact number and the prestress no longer holds, as shown in Fig.~\ref{fig:fig1}.
This suggests the possibility that the non-Debye scaling of the vDOS is a robust feature of amorphous solids, irrespective of the jamming transition.
We will further discuss this point in Sec.~\ref{discussion}. 
From the data for $e\geq0.1$, we can appreciate the full frequency dependence of the non-Debye contribution to the density of states and estimate where it begins and where it ends. Thus, in the following, we focus on the case $e\geq0.1$ to discuss the dimensional dependence of the frequency range where the non-Debye scaling holds. 

\begin{figure*}
    \centering
    \includegraphics{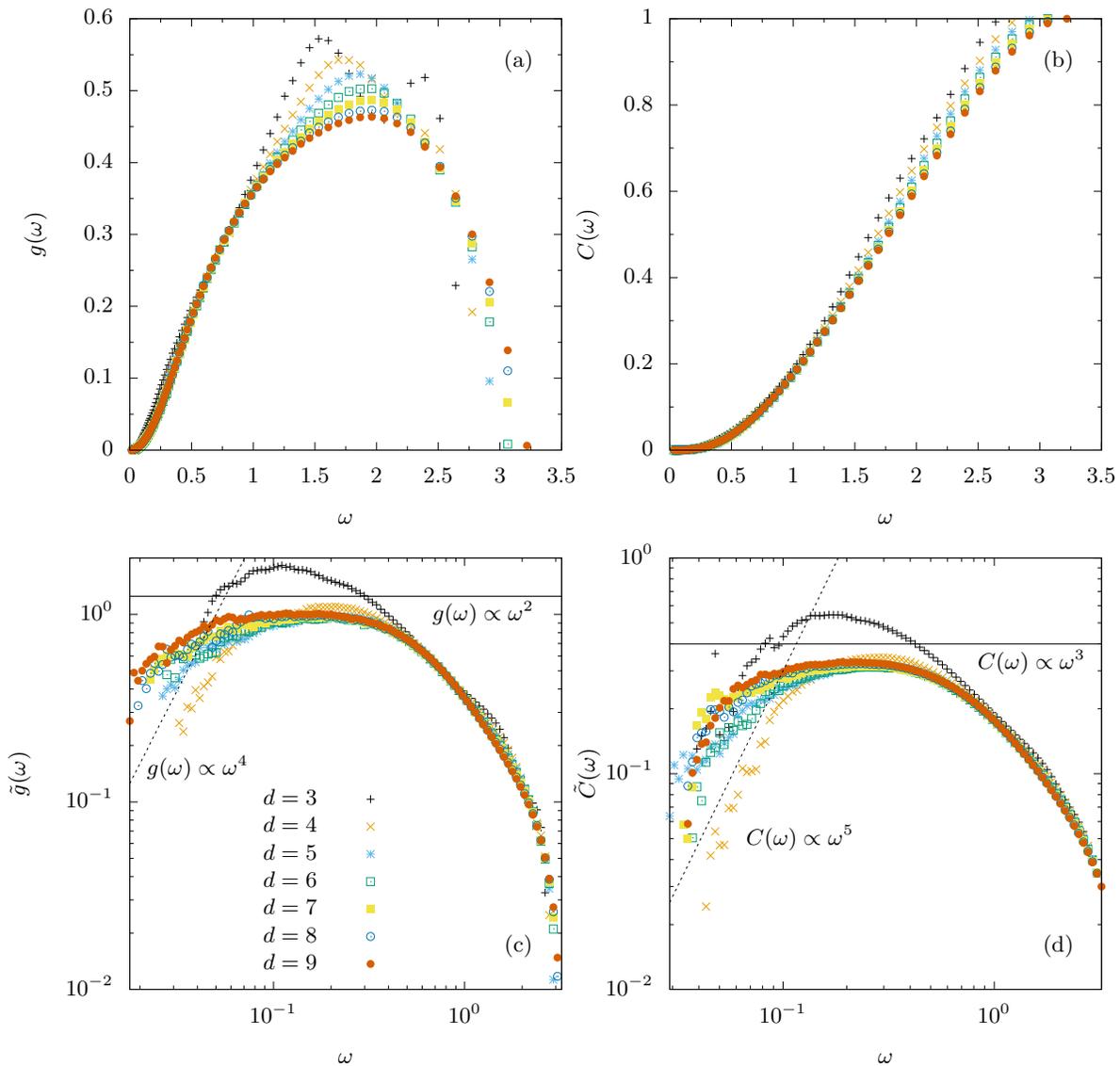}
\caption{
(a) The vDOS, (b) the CD, (c) the reduced vDOS, and (d) the reduced CD for $e=0.1$. For the data for the other prestresses $e$, see Appendix~\ref{sec:appendix}.
The solid lines in (c) and (d) are the frequency dependence of the non-Debye scaling. The dotted lines in (c) and (d) have slope $2$ and $3$, respectively, indicating the frequency dependence of the QLVs, $g(\omega)\propto \omega^4$.
}
    \label{fig:fig4}
\end{figure*}

\begin{figure}
    \centering
    \includegraphics[width=0.45\textwidth]{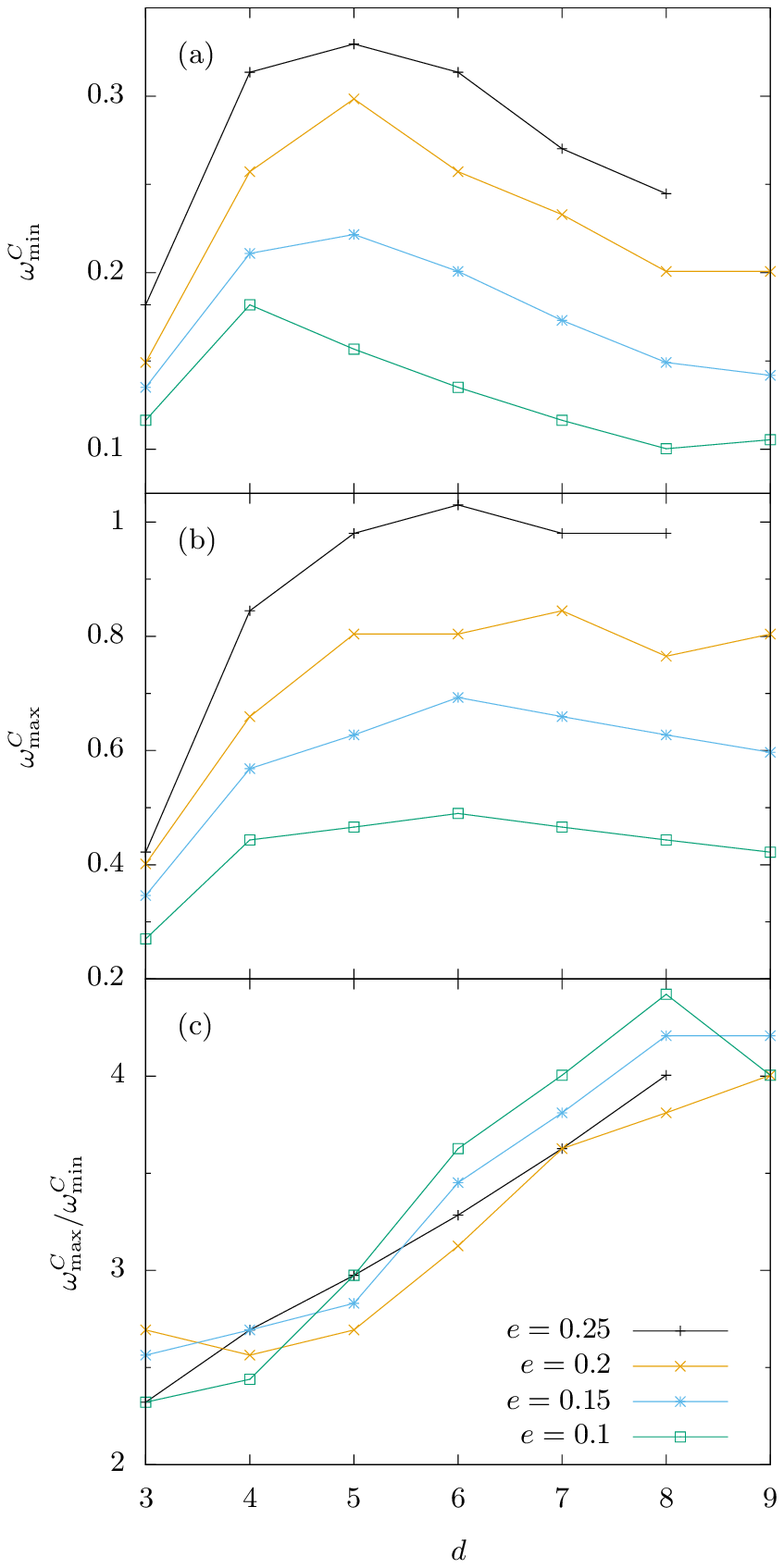}
\caption{
(a) $\omega_{\mathrm{min}}^C$, (b) $\omega_{\mathrm{max}}^C$, and (c) $\omega_{\mathrm{max}}^C/\omega_{\mathrm{min}}^C$ as functions of $d$.
The data are connected by lines (this is not a fit).
For the data for $\omega_{\mathrm{min}}^g$ and $\omega_{\mathrm{max}}^g$, see 
Appendix~\ref{sec:appendix}.}
    \label{fig:fig5}
\end{figure}

In Fig.~\ref{fig:fig4}, we offer several data representations of the 
density of states for various numbers of dimensions at a fixed prestress of $e=0.1$. The data for the other values of the prestresses $e$ are available in Appendix~\ref{sec:appendix}.

Figures~\ref{fig:fig4}(a) and (b) show the vDOS and the CD, respectively.
These plots indicate that as $d$ increases at a fixed $e$, the vDOS and the CD overall converge to dimension-independent functions. This finding is consistent with the results for the scaling behaviors of $\delta z/2d$ and $\omega^\ast$ in Fig.~\ref{fig:fig1}. We note that the two peaks of the vDOS at $\omega \sim 1.5$ and $2.5$ for $d=3$ disappear and merge to form a single broad peak in a large number of dimensions.

To examine the non-Debye scaling in the low-frequency regime, we plot the reduced versions of these functions for the same prestress, $e=0.1$, in Figs.~\ref{fig:fig4}(c) and (d).
We can clearly see that these functions depend on the dimensionality in the lowest-frequency region.
The results in $d \ge 4$ nearly collapse for $\omega \gtrsim 0.1$, and a fit to the data to a plateau corresponding to the non-Debye scaling is convincing, even on a logarithmic scale.
This implies that the prefactor of the quadratic non-Debye scaling does not depend on $d$ for $d \ge 4$, and is solely controlled by the value of prestress $e$.

On the other hand, when we focus on $\omega \lesssim 0.1$, we see that the non-Debye scaling region extends to lower frequencies with increasing $d$ from  $d = 4$ to $9$.
Since the quartic frequency dependence of the QLVs $g_{QLV}(\omega)\sim\omega^4$ has been reported in a previous study~\cite{Kapteijns2018Universal}, we show the corresponding dotted lines of slope $2$ in Fig.~\ref{fig:fig4}(c) and of slope $3$ in Fig.~\ref{fig:fig4}(d). These fits suggest that the QLVs survive up to dimension $d=9$, but appear at lower frequencies for larger $d$. 

To quantitatively study this behavior, we measured the frequency width of the non-Debye scaling region.
We extracted the two frequencies at which the reduced distribution is smaller than its maximum by 10\%,
which we denote by $\omega_{\mathrm{max}}$ and $\omega_{\mathrm{min}}$ (with the convention $\omega_{\mathrm{min}}<\omega_{\mathrm{max}})$.
We use a superscript $g$ or $C$ to specify the function from which each of these frequencies was extracted, i.e., four frequencies are considered for each $e$: $\omega_{\mathrm{min}}^g$, $\omega_{\mathrm{max}}^g$, $\omega_{\mathrm{min}}^C$, and $\omega_{\mathrm{max}}^C$.
In Fig.~\ref{fig:fig5}, we plot (a) $\omega_{\mathrm{min}}^C$, (b) $\omega_{\mathrm{max}}^C$, and (c) $\omega_{\mathrm{max}}^C/\omega_{\mathrm{max}}^C$ as functions of $d$ for $e\geq0.1$.
The data for $\omega_{\mathrm{min}}^g$ and $\omega_{\mathrm{max}}^g$, which exhibit qualitatively the same behaviors as $\omega_{\mathrm{min}}^C$ and $\omega_{\mathrm{max}}^C$, are shown in Appendix~\ref{sec:appendix}.
The value of $\omega_{\mathrm{min}}^C$ decreases with increasing $d$ for $d\gtrsim4$, whereas $\omega_{\mathrm{max}}^C$ increases at small $d$ and then quickly saturates for $d\gtrsim5$.
These results are consistent with the observations in Figs.~\ref{fig:fig4}(b) and (d).
From these two results, we conclude that the non-Debye scaling region applies over a broader frequency range with increasing $d$.
By dividing $\omega_{\mathrm{min}}^C$ by $\omega_{\mathrm{max}}^C$, we clarify this tendency in Fig.~\ref{fig:fig5}(c).
This plot shows that $\omega_{\mathrm{max}}^C/\omega_{\mathrm{min}}^C$ increases for all $e$ as the number of spatial dimensions increases.
Therefore, we conclude that the non-Debye scaling region becomes broader for larger dimensionality. 

Although our data are limited to $d \leq 9$, the non-Debye scaling region continuously extends with increasing dimensionality without any sign of saturation; thus, we expect that the vDOS of a jammed particle system approaches the gapless non-Debye scaling in the large-dimensional limit; namely, it converges to the form predicted by effective medium theory~\cite{Degiuli2014Effects} and by replica theory for a perceptron~\cite{Franz2015Universal}. We are not aware of any theoretical prediction for how fast the large $d$ limit should be reached by increasing $d$, but the data presented in this work suggest that the convergence, even if real, is rather modest as the frequency width of the non-Debye scaling seems to grow linearly with $d$. Similar convergences towards the large $d$ limit in the context of mean-field theory is not infrequent~\cite{PhysRevLett.104.255704,doi:10.1146/annurev}. 


\section{Summary and discussion}

\label{discussion} 

In this work, we have numerically studied the low-frequency vibrational properties of jammed particles in $d=3$--$9$ spatial dimensions.
We first showed that the prestress $e = (d-1)\left<1/r_{ij}-1\right>_{ij}$ is an appropriate control parameter for studying jamming scaling behaviors in different dimensions.
In particular, the excess contact number divided by $2d$, $\delta z/2d$, and the onset frequency of the flat region of the vDOS, $\omega^\ast$, in various dimensions were shown to follow universal functions of the prestress $e$; near the jamming transition, $\delta z/2d \approx  1.8 e^{1/2}$ and $\omega^\ast \approx  1.95 e^{1/2}$ work almost perfectly in any number of dimensions.
Then, by comparing the vDOS in different dimensions at the same prestress $e$, we studied the dimensional dependence of the vDOS in the low-frequency region.
Our system sizes of $N=8000$--$128000$ enabled us to capture the full frequency range of the non-Debye scaling $g(\omega) \sim \omega^2$ in $d=3$--$9$.
We found that the non-Debye scaling appears below $\omega^\ast$ in all dimensions and that the frequency width of the non-Debye scaling region grows with increasing dimensionality without any sign of saturation.
From these findings, we expect that the vDOS of a real packing of particles converges to the gapless non-Debye scaling in the large-dimensional limit, thus fully supporting the prediction of effective medium theory~\cite{Degiuli2014Effects} and replica theory for a perceptron~\cite{Franz2015Universal}.

Related to this finding, two comments are in order.
The first concerns the precise form of the dimensional dependence of $\omega_{\mathrm{max}}^C/\omega_{\mathrm{min}}^C$.
A packing of particles in finite dimensions will include rattler particles that do not contribute to the rigidity~\cite{O'Hern2003Jamming}.
The presence of rattlers is a kind of finite-dimensional effect, and previous studies have established that the fraction of rattlers decreases with increasing dimensionality~\cite{Charbonneau2012}.
This decrease is very rapid, with the fraction following $\propto e^{- \alpha d}$ with a constant $\alpha$~\cite{Charbonneau2012}.
Based on this observation, one might expect that the finite-dimensional effect in the vDOS should also vanish exponentially with increasing $d$, i.e., that $\omega_{\mathrm{max}}^C/\omega_{\mathrm{min}}^C$ should increase exponentially.
However, we found that the dimensional dependence of $\omega_{\mathrm{max}}^C/\omega_{\mathrm{min}}^C$ is not very dramatic, at least in $d \leq 9$, and that the data are still compatible with a linear dependence on $d$.
It would be interesting to determine whether $\omega_{\mathrm{max}}^C/\omega_{\mathrm{min}}^C$ ultimately grows exponentially at $d \geq 10$, although the computational cost of such a study is beyond our reach for the moment.

Second, our study established that the non-Debye scaling holds even far from the jamming transition point, as shown in Fig.~\ref{fig:fig3} and discussed in the corresponding paragraph.
This result suggests that the non-Debye scaling may be more universal than discussed so far in the context of the jamming transition.
In fact, not only the theories for jammed solids~\cite{Franz2015Universal,Degiuli2014Effects,Degiuli2014Force} but also elasticity theory with a fluctuating elastic modulus~\cite{Schirmacher2006Thermal,Schirmacher2007Acoustic} predict a quadratic frequency dependence of the vDOS near the BP frequency.
The latter theory is not rooted in jammed materials and regards glasses as elastic continua with a spatially fluctuating elastic modulus to describe the universal behaviors of the low-frequency excitations~\cite{Schirmacher2006Thermal,Schirmacher2007Acoustic}.
In this respect, it will be interesting to study whether amorphous solids with other potentials, such as the Lennard-Jones potential, also exhibit non-Debye scaling in large dimensions.
This topic will be addressed in future work.





\begin{acknowledgments}
This work was supported by JSPS KAKENHI Grant Numbers 19J20036, 17K14369, 17H04853, 16H04034, 18H05225, 19K14670, and 19H01812.
This work was also partially supported by the Asahi Glass Foundation.
The research leading to these results has received
funding from the Simons Foundation (Grant No. 454933, Ludovic
Berthier).
\end{acknowledgments}

\onecolumngrid

\clearpage
\appendix

\section{Additional data for the vDOS and the CD}\label{sec:appendix}

We report additional data for the vDOS and the CD with different $e$ values (supplementing Fig.~\ref{fig:fig4}) and for $\omega_{\mathrm{min}}^g$ and $\omega_{\mathrm{max}}^g$ (supplementing Fig.~\ref{fig:fig5}).

\begin{figure}[h]
    \centering
    \includegraphics[width=0.9\textwidth]{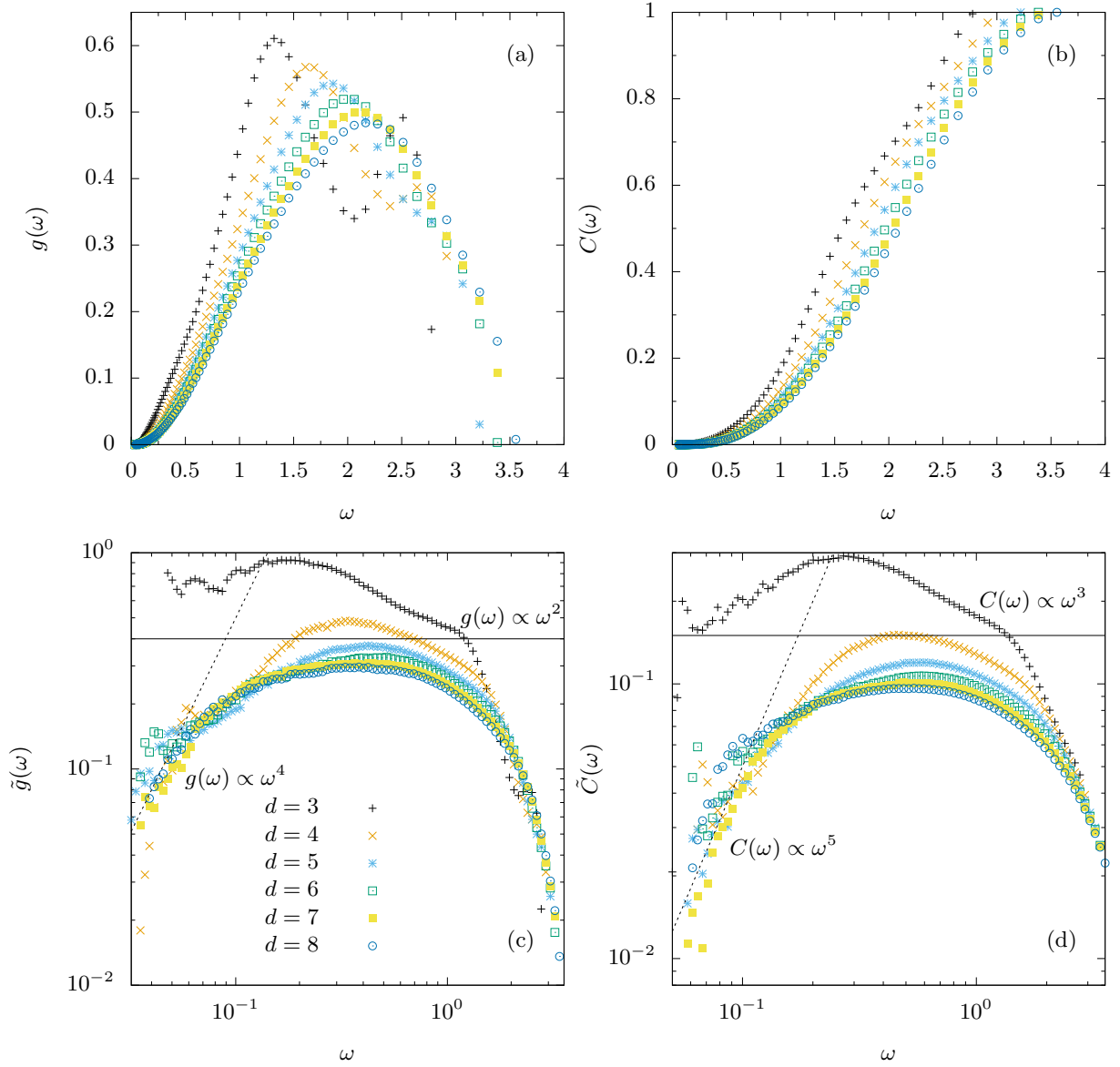}
    \caption{
(a) The vDOS, (b) the CD, (c) the reduced vDOS, and (d) the reduced CD for $e=0.25$.
Due to numerical limitations, we could not calculate the results for $d=9$.
The solid lines in (c) and (d) are the frequency dependence of the non-Debye scaling.
The dotted lines in (c) and (d) have slope $2$ and $3$, respectively, indicating the frequency dependence of the QLVs.
}
    \label{fig:fig6}
\end{figure}

\begin{figure}
    \centering
    \includegraphics[width=0.9\textwidth]{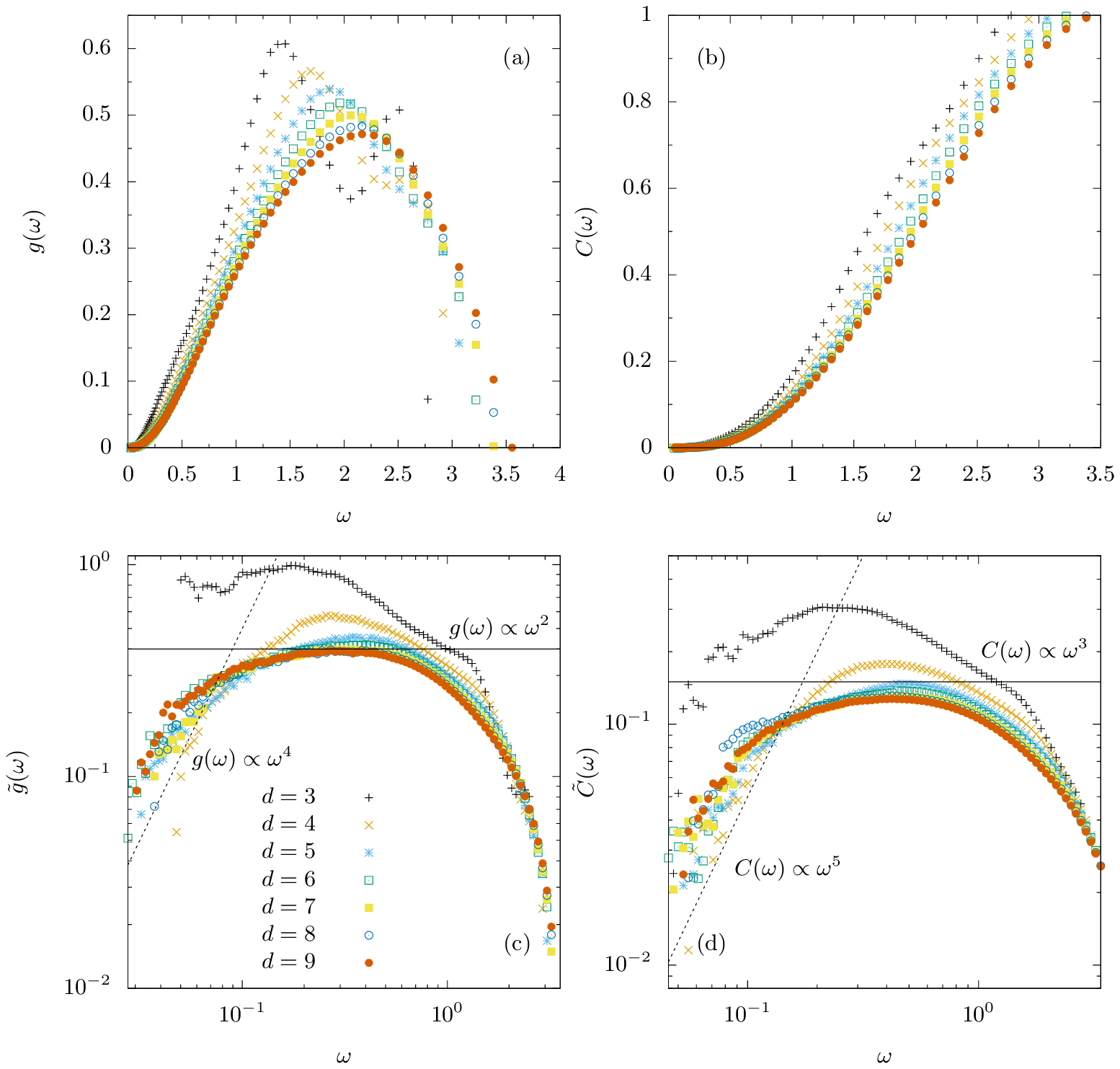}
\caption{
(a) The vDOS, (b) the CD, (c) the reduced vDOS, and (d) the reduced CD for $e=0.2$.
The solid lines in (c) and (d) are the frequency dependence of the non-Debye scaling.
The dotted lines in (c) and (d) have slope $2$ and $3$, respectively, indicating the frequency dependence of the QLVs.
}
    \label{fig:fig7}
\end{figure}

\begin{figure}
    \centering
    \includegraphics[width=0.9\textwidth]{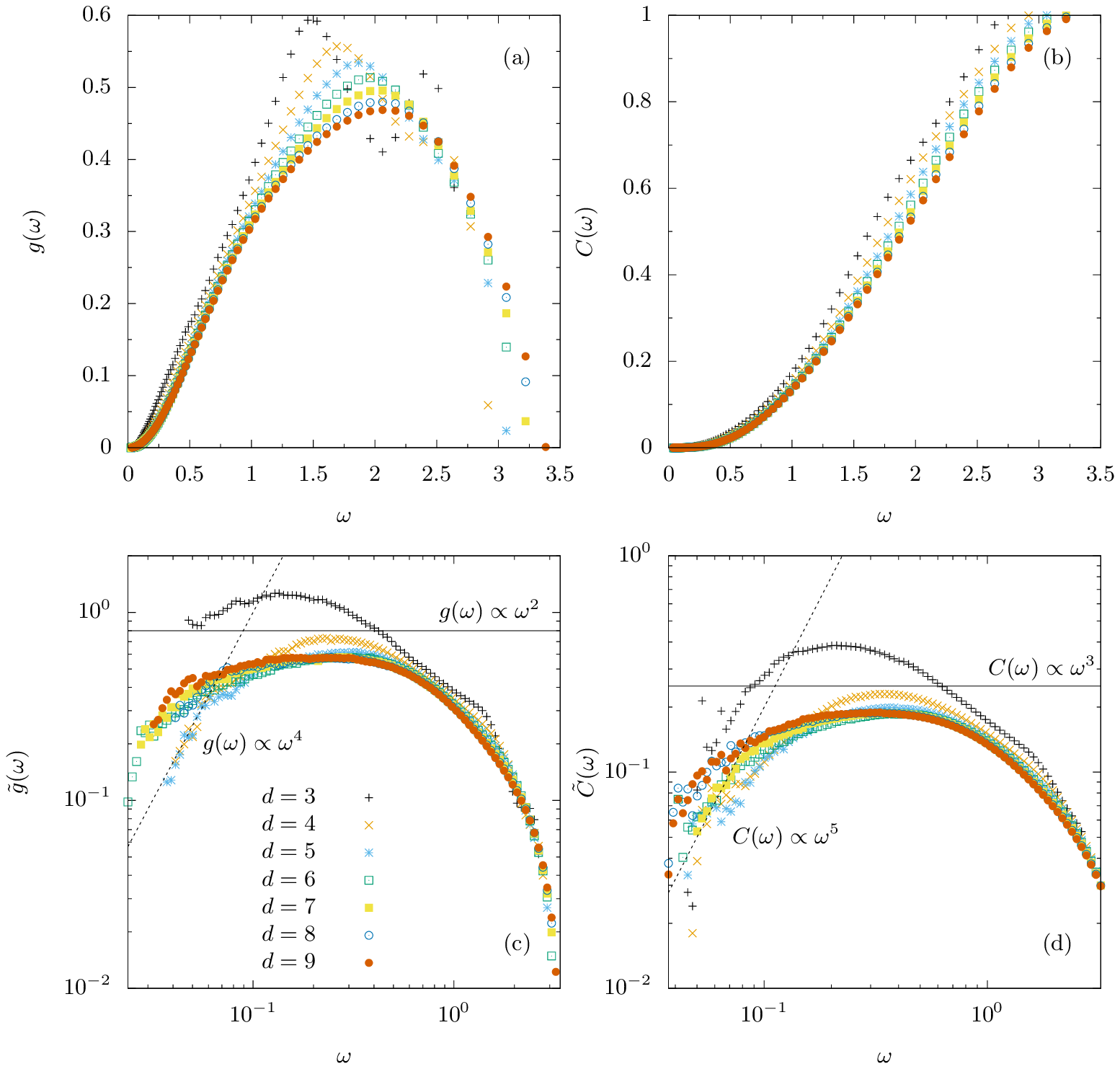}
\caption{
(a) The vDOS, (b) the CD, (c) the reduced vDOS, and (d) the reduced CD for $e=0.15$.
The solid lines in (c) and (d) are the frequency dependence of the non-Debye scaling.
The dotted lines in (c) and (d) have slope $2$ and $3$, respectively, indicating the frequency dependence of the QLVs.
}
    \label{fig:fig8}
\end{figure}

\begin{figure}
    \centering
    \includegraphics[width=0.9\textwidth]{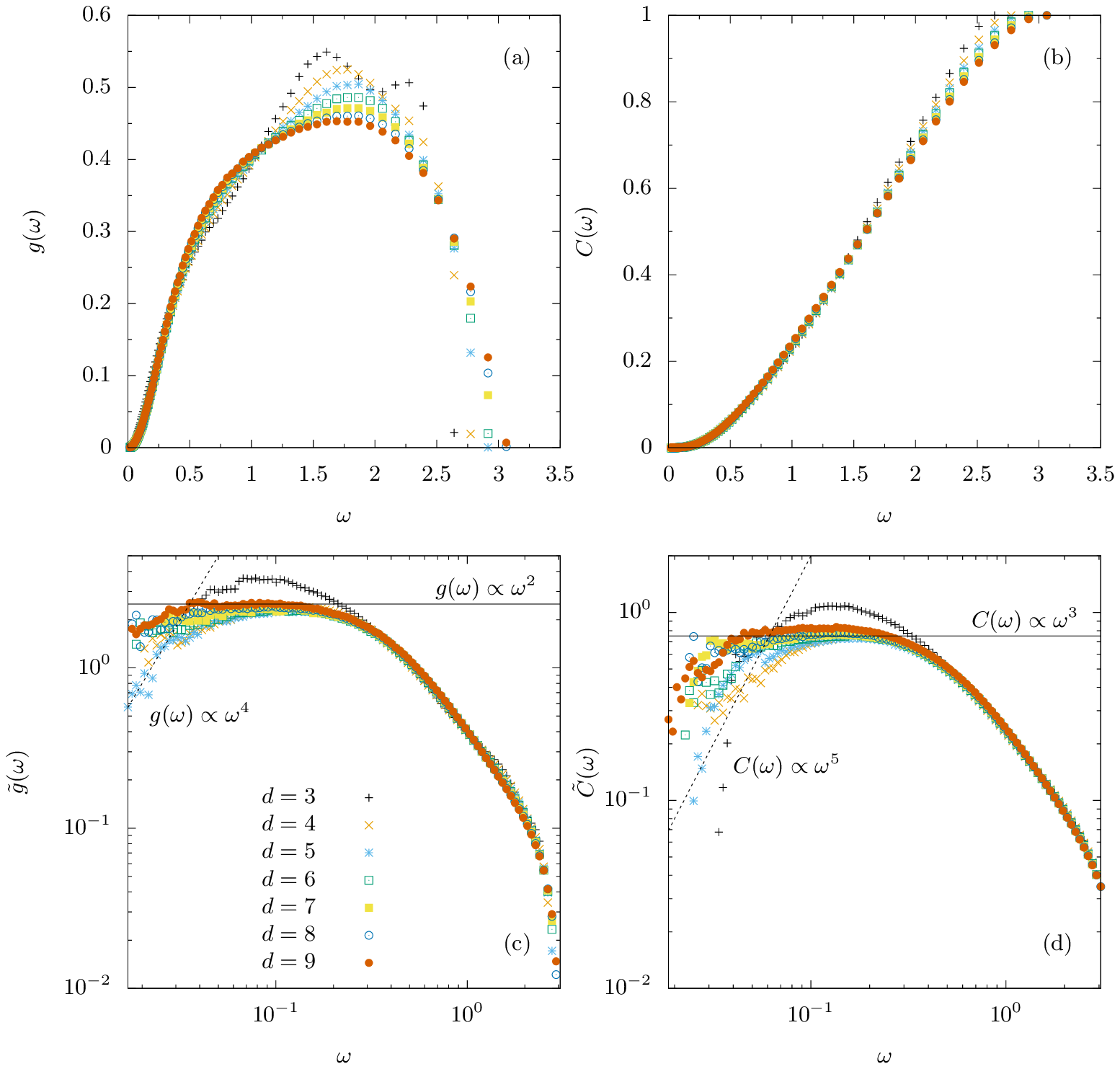}
\caption{
(a) The vDOS, (b) the CD, (c) the reduced vDOS, and (d) the reduced CD for $e=0.05$.
The solid lines in (c) and (d) are the frequency dependence of the non-Debye scaling.
The dotted lines in (c) and (d) have slope $2$ and $3$, respectively, indicating the frequency dependence of the QLVs.
}
    \label{fig:fig9}
\end{figure}

\begin{figure}
    \centering
    \includegraphics[width=0.9\textwidth]{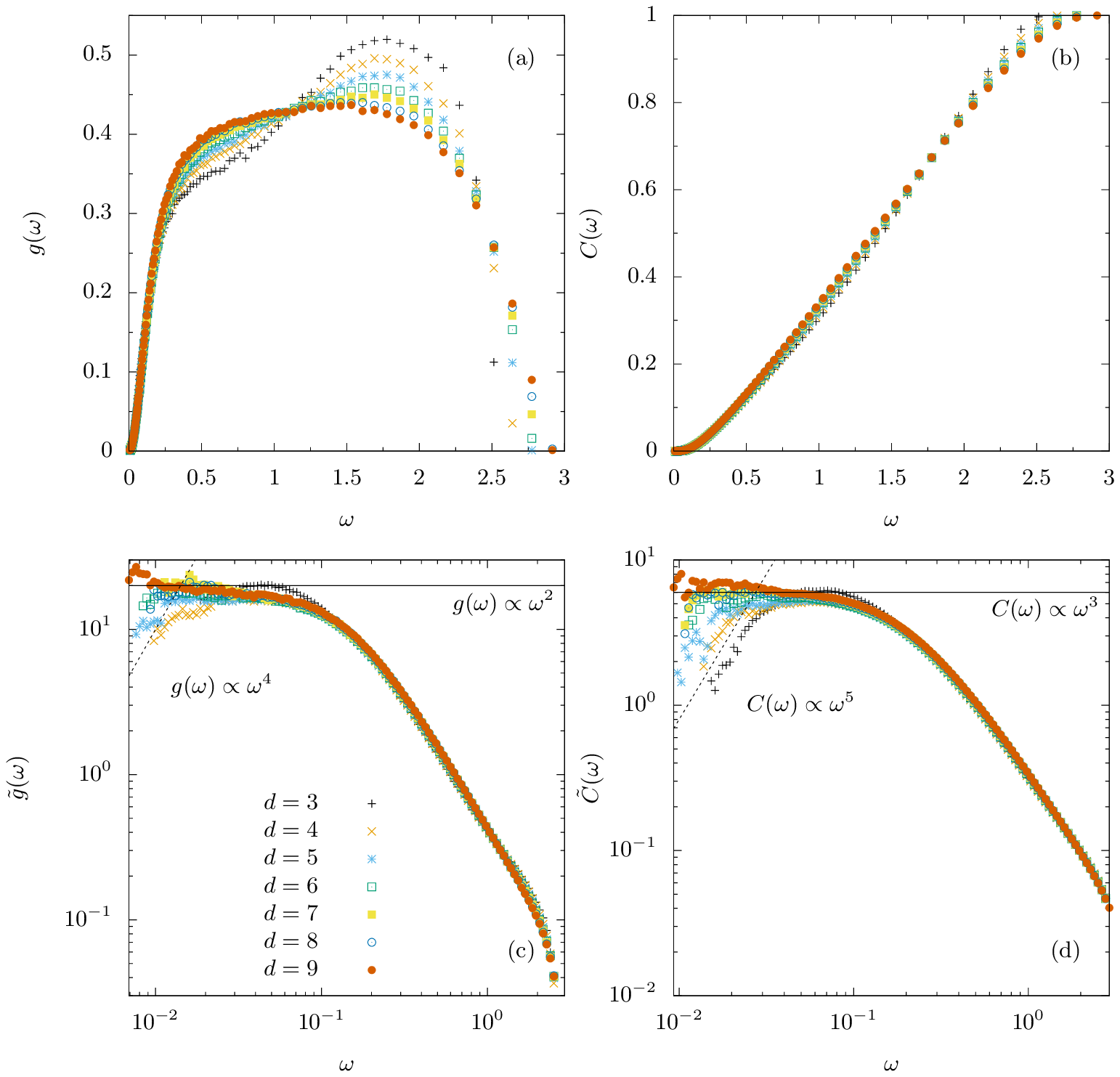}
\caption{
(a) The vDOS, (b) the CD, (c) the reduced vDOS, and (d) the reduced CD for $e=0.01$.
The solid lines in (c) and (d) are the frequency dependence of the non-Debye scaling.
The dotted lines in (c) and (d) have slope $2$ and $3$, respectively, indicating the frequency dependence of the QLVs.
}
    \label{fig:fig10}
\end{figure}

\begin{figure}
    \centering
    \includegraphics{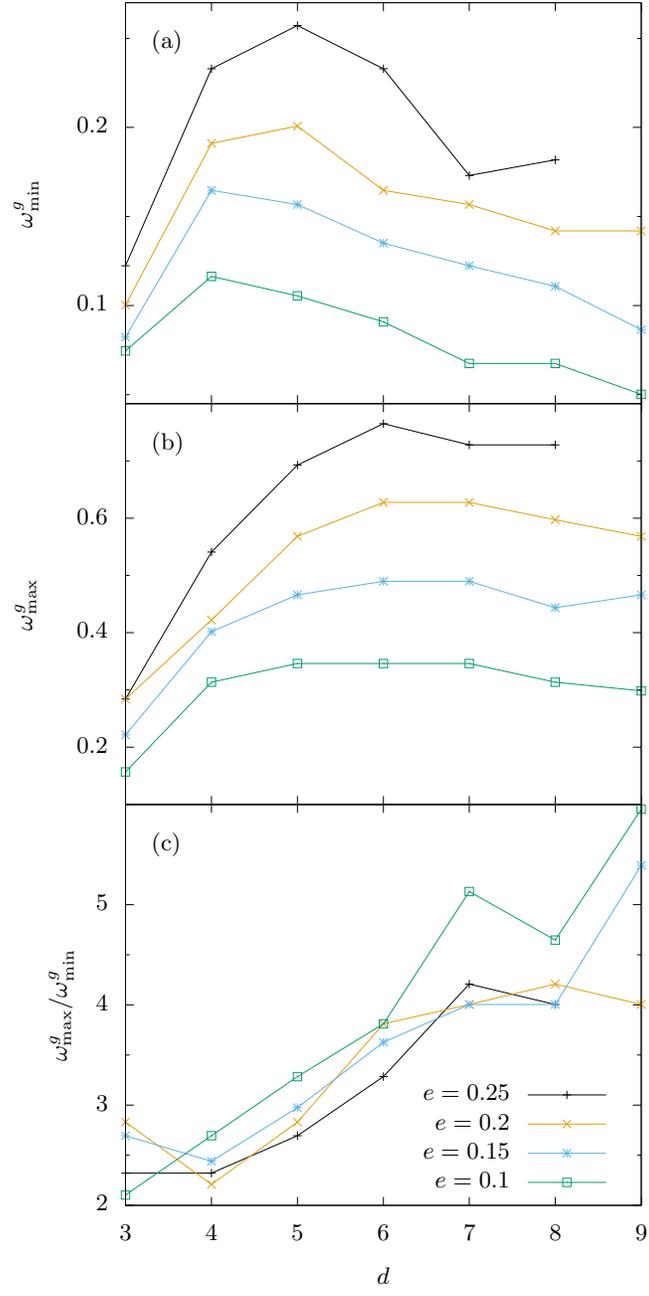}
\caption{
(a) $\omega_{\mathrm{min}}^g$, (b) $\omega_{\mathrm{max}}^g$, and (c) $\omega_{\mathrm{max}}^g/\omega_{\mathrm{min}}^g$ as functions of $d$.
The lines are simple guides for the eye.
}
    \label{fig:fig11}
\end{figure}

\clearpage
\twocolumngrid

\bibliographystyle{unsrt}
\bibliography{manuscript}

\end{document}